# A model of crack based on dislocations in smectic A liquid crystals

Tian You Fan, Zhi Yi Tang

School of Physics, Beijing Institute of Technology

Beijing 100081, China

**Abstract** A plastic crack model for smectic A liquid crystals under longitudinal shear is suggested. The solution of screw dislocation in smectic A is the key in which the correct result is just obtained by overcoming a longstanding puzzle [19]. We further use the dislocation pile-up principle and the singular integral equation method, construct the solution of the crack in the phase. From the solution we can determine the size of the plastic zone at the crack tip and the crack tip opening (tearing) displacement, which are parameters being relevant to the local stability/instability of materials. The results may be useful in developing soft-matter mechanics.

It is well-known that liquid crystals and quasicrystals are the important phases in condensed matter, and fascinating from a fundamental point of view. Liquid crystals are also significant on account of a number of current and potential applications. They belong，macroscopically, to an intermediate phase between isotropic liquid and solid. The mechanical properties of the phase are very interesting and have been extensively studied, see de Gennes and Prost [1], Kleman and Oswald [2], Oswald and Pieranski [3] for example. The need of basic research and engineering applications requires promoting the study further, especially of three-dimensional elasticity, plasticity, defects and dynamics. Among defects, dislocations and focal conics have been investigated, see for example the above references [1-3], Landau and Lishitz [4], Fujii et al [5]. In addition, Brostow et al [6] have carried out numerical simulation on crack formation and propagation in polymer liquid crystals. Because liquid crystals including smectic ones can be classified as monomer liquid crystals (MLCs) irrespectively of the fact whether they can or cannot polymerize and polymer liquid crystals (PLCs). That classification is due to Samuski [7] and has been used by a number of authors [8, 9]. The present model applies to MLCs, smectic A phase in particular. To develop the pioneering work of Brostow et al [6], Fan [10] studied the novel defect in liquid crystal very recently. In the respect of dynamics, the hydrodynamics, elastodynamics and wave propagation were discussed in the papers and books of de Gennes and Prost [1], de Gennes and Kleman [11], Oswald and Pieranski [3] and Landau and Lishitz [4]. In addition, on the plasticity of liquid crystals, de Gennes and Prost [1], Oswald and Pieranski [3] have done the pioneering work. However the plastic theory of liquid crystals has not been developed so far, either microscopically or macroscopically. This leads to some difficulties on the study of plasticity of liquid crystals. On the three-dimensional elasticity for liquid crystals, there is a preliminary work, see for example Fan [12], but more difficult topics have not yet been engaged. Owing to the difficulty mentioned above, we suggest a phenomenological model to conduct a plasticity problem for smectic liquid crystals. An analytic solution of plasticity-crack interaction are presented, which may provide a basis for fundamental research in soft matter physics and chemistry.

It is well-known that there are inherent connections between dislocations and cracks, for example, piling-up of dislocations can result in a crack, which may strongly influence the mechanical behaviour of liquid crystals. Cracks affect the plasticity of liquid crystals, but the mechanism is not so clear. A coupling between fracture and plasticity makes the problem more complicated. Fan [10] discussed a plastic crack in a smectic liquid based on the dislocation pile-up model. This model has been used for crystals by Bilby et at [12, 13], and for quasicrystals by Fan et al [14, 15]. But the work carried out in Ref [10] contains some approximations, in which the movement of dislocations do not strictly follow the Peach-Koehler force rule, the reason for this lies in the limitation of methodology adopted. In Ref [10] work conducted is edge dislocation pile-up model. The present work takes screw dislocation pile-up model in which the movement of the dislocations can strictly obey the Peach-Koehler force rule.

Consider a smectics A liquid crystal whose layers are in the $xy$–plane, and a crack dislocation group with length $2l$ along $x$–axis subjected to a uniform shear stress $\sigma_{yz} = \tau^{(\infty)}$ shown in Fig.1, in which the single screw dislocation has Burgers vector $b = (0,0,b)$. Because the deformation is assumed to be independent from variable $y$, the figure depicts any transverse section of the body. At the crack tip there is an screw dislocation pile-up with length $d$, whose value is unknown and to be determined. We call the pile-up as sliding dislocation group. Within the zone defined by $y = 0, l \leq |x| \leq l + d$, a counter direction shear stress $\tau_c$ is applied, the value of which represents the yield strength of the materials macroscopically. In other words the dislocation pile-up zone is the plastic zone. The physical meaning of $\tau_c$ can be referred to the monograph [1] (p.499).

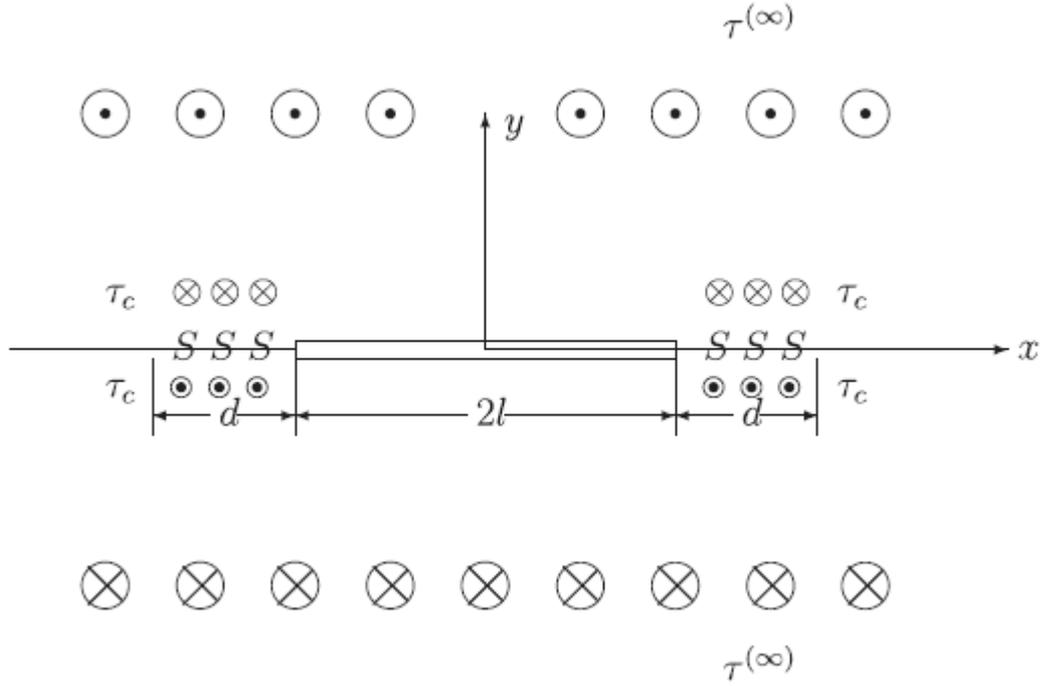

Fig.1    A plastic crack in a smectics A under longitudinal shear

The model can be formulated by the following (equivalent) boundary conditions:

$$\left.\begin{array}{l} (x^2+y^2)^{1/2} \to \infty: \quad \sigma_{ij}=0 \\ y=0, \ |x|<l: \quad \sigma_{yz}=-\tau^{(\infty)} \\ y=0, \ l<|x|<l+d: \sigma_{yz}=-\tau^{(\infty)}+\tau_c \end{array}\right\} \quad (1)$$

A quantitative description of the problem needs to introduce relevant basic governing equations. The mathematical model (1) requires the non-linear (plastic) problem to be linearized although it contains a unknown constant $d$. This greatly simplifies the mathematical solution, and allows us to choose linear governing equations.

According to de Gennes and Prost [1] or Landau and Lifshitz [4] the stress tensor for smectic liquid crystals (exactly speaking for smectics A)

is

$$\left.\begin{array}{l}\sigma_{xx}=\sigma_{yy}=K_1\nabla^2\dfrac{\partial u}{\partial z}\\[4pt]\sigma_{zz}=\rho_0 B^{'}\dfrac{\partial u}{\partial z}\\[4pt]\sigma_{zx}=\sigma_{xz}=-K_1\nabla^2\dfrac{\partial u}{\partial x}\\[4pt]\sigma_{zy}=\sigma_{yz}=-K_1\nabla^2\dfrac{\partial u}{\partial y}\\[4pt]\sigma_{xy}=\sigma_{yx}=0\end{array}\right\} \qquad (2)$$

in which $u\equiv u_z$ is the displacement of $z$–component, $B^{'}=B-C^2/A$, $\rho_0$ and $K_1$ are materials constants, see [4]. Physically the constant $B^{'}$ describes material modulus relating to deformation arising from displacements, while $K_1$ describes material modulus relating to deformation arising from curvature. It is evident that to the stress tensor (2) only the symmetry part, according to the Landau's terminology, is considered, the viscous part is omitted here by following the methodology adopted in the liquid crystal study community. Substituting equations (2) into the equilibrium equations

$$\dfrac{\partial \sigma_{ij}}{\partial x_j}=0 \qquad (3)$$

yields the final governing equation

$$\nabla^2\nabla^2 u=0 \qquad (4)$$

where $\nabla^2=\dfrac{\partial^2}{\partial x^2}+\dfrac{\partial^2}{\partial y^2}$. Equation (4), which can also be derived from the energy minimization, is identical to those given by de Gennes and Prost [1], Oswald and Pieranski [3] and Landau and Lishitz [4].

The solution of the plastic crack is reduced to solve equation (4) under boundary conditions (1), which can be called the 'boundary value problem (4),(1)'.

To solve the 'boundary value problem (4),(1)' we can refer to the experiences in crystals[13,14], quasicrystals [15,16], we can also refer to the experience in liquid crystals for edge dislocation more recently, i.e., to transform the proble into the following singular integral equations

$$\int_L \frac{f(\xi)d\xi}{\xi - x} = \frac{\tau(x)}{A} \quad (5)$$

in which $f(\xi)$ is a dislocation density function, $\xi$ the dislocation source point coordinate, and $x$ the field point coordinate on the real axis, $L$ represents interval $(-(l+d), l+d)$, and $\tau(x)$ the shear stress distribution at the region $z = 0, |x| \leq l + d$, i.e.,

$$\tau(x) = \begin{cases} -\tau^{(\infty)}, & |x| < l \\ -\tau^{(\infty)} + \tau_c, & l < |x| < l+d \end{cases} \quad (6)$$

and the key is constant $A$ how can be chosen? This depends upon the solution of screw dislocation of smectic A liquid crystals. Kleman [17] and Pershan [18] have derived the solution for a long time, but it is wrong, this results in a longstanding puzzle in study, which can named de-Gennes-Kleman-Pershan paradox.

Fan and Li [19] overcome the puzzle, and gave a correct solution.

$$A = \left(\frac{b}{\pi}\right) K_1 D_1, \quad D_1 = \frac{-\frac{8}{3}\pi\alpha}{(R_0 + r_0)\left(\frac{\pi}{4}\alpha\beta + \frac{b^2 K_1}{8\pi}\right)\ln\frac{R_0}{r_0} + \frac{\pi}{320}\alpha\gamma(R_0 - r_0)} \tag{7}$$

in which

$$\left. \begin{array}{l} \alpha = \left(\dfrac{b}{2\pi}\right)^4 \rho_0 B' \\[6pt] \beta = 2 + \dfrac{32\pi^2}{3} \\[6pt] \gamma = 75 - 160\pi^2 + 256\pi^4 \end{array} \right\} \tag{8}$$

(Note that the constant $A$ here should not be confused with the material constant $A$ in equation (2)). In terms of the singular integral equation theory of Muskhelishvili [18] (p.251), the integral equations (5) under condition (6) has the solution

$$\begin{aligned} f(x) &= -\frac{1}{\pi^2 A}\sqrt{\frac{x+(l+d)}{x-(l+d)}}\int_L \sqrt{\frac{\xi-(l+d)}{\xi+(L+d)}}\tau(\xi)\frac{d\xi}{\xi-x} \\ &= -\frac{1}{\pi^2 A}\sqrt{\frac{x+(l+d)}{x-(l+d)}}\left\{i\left[2\tau_c \cos^{-1}\left(\frac{l}{l+d}\right) - \tau^{(\infty)}\pi\right]\right\} \\ &\quad + \frac{\tau_c}{\pi^2 A}\left[\cosh^{-1}\left|\frac{(l+d)^2 - lx}{(l+d)(l-x)}\right| - \cosh^{-1}\left|\frac{(l+d)^2 + lx}{(l+d)(l+x)}\right|\right] \end{aligned} \tag{9}$$

(the details of the mathematical calculation are quite lengthy and are omitted here), in which $i = \sqrt{-1}$, and $A$ is defined by equation (7). Because the dislocation density $f(x)$ should be a real function, the factor multiplying the imaginary number $i$ in the first term of right-hand side of formula (8) must be zero, this leads to

$$2\tau_c \cos^{-1}\left(\frac{l}{l+d}\right) - \tau^{(\infty)}\pi = 0$$

i.e.,

$$d = l\left[\sec\left(\frac{\pi\tau^{(\infty)}}{2\tau_c}\right) - 1\right] \quad (10)$$

This determines the plastic zone size.

From solution (9) we evaluate amount of dislocations $N(x)$ such as

$$N(x) = \int_0^x f(\xi)d\xi \quad (11)$$

Substituting (9) (coupled with (10)) into (11) we can get values of $N(l+d)$ and $N(l)$, so the amount of dislocation motion is

$$\delta = b[N(l+d) - N(l)] = \frac{2bl\tau_c}{\pi^2 A}(\ln\frac{l+d}{l}) = \frac{2\tau_c l}{\pi\rho_0 B}\ln\sec(\frac{\pi\tau^{(\infty)}}{2\tau_c}) \quad (12)$$

This is the crack tip opening (tearing) displacement, which is an important parameter.

We suggest the following fracture criterion

$$\delta = \delta_c \quad (13)$$

which can be used for determining the thermodynamic stability/instability of the material, $\delta_c$ is the critical value of the crack tip opening displacement, which can be measured by experiments, and is a material constant of the liquid crystals. The equation (13) describes a critical state of equilibrium of the plastic crack. When $\delta < \delta_c$, the crack does not propagate, but when $\delta > \delta_c$, the crack will propagate. By using this criterion, the limiting value of the applied stress $\tau^{(\infty)}$ or the limit value of the crack size $l$ can be determined.

The above treatment is a macro-description (or the continuum model), but a micro-description (or a micro-mechanism) can be given as follows. By introducing the de Gennes theory (refer to [1]), the yield stress is

$$\tau_c \sim \frac{\pi \gamma^2}{a_0 k_B T \ln(v_0/v_1)} \qquad (14)$$

where

$$\gamma \approx \frac{1}{2}\sqrt{K_1 B a_0^2}/\eta, \quad \eta \sim \sqrt{\frac{K_1}{B}}, \quad v_0 = 10^{33}\,\text{s}^{-1}\text{cm}^{-3}, \quad v_1 = 1\,\text{s}^{-1}\text{cm}^{-3} \qquad (15)$$

and $a_0$ represents the thickness of the layer of the smectic, whose value is almost equivalent to the magnitude of a Burgers vector ($10\,\text{nm}$), $k_B$ is the Boltzman constant, $T$ the absolute temperature, $v_0$ and $v_1$ the fluctuation frequencies, respectively.

Substituting expression (14) into equations (10) and (12) respectively one reveals the physical sense of the plastic zone size (or dislocation sliding width) and crack tip opening displacement (or amount of dislocation motion) in-depth. Owing to the limitation of space a detailed discussion is not given here.

Crack and plasticity are difficult topics in liquid crystals. One of reasons for this lies in there being lack no theory of plasticity, at least, there is an absence of macroscopic plastic constitutive equation so far. Here we have adopted a phenomenological model to discuss the problem. In this way we obtain some physical quantities for describing the coupling between fracture and plasticity. The methodology developed here is generally effective for other problems in smectics and other classes of liquid crystals. Results obtained by the present author and co-workers will be reported in other papers, e.g. [19].

The model suggested in Ref [10] has its limitation itself, for example, the Peach-Koehler force for the configuration is not strictly along the x-direction, only has a component along the direction. This means the movement of dislocations is not strictly along the x-direction, so the dislocation pile-up. So that the assumption of the movement of dislocations along the x-direction is only approximate, the assumption of the dislocation pile-up along the direction is also approximate. For the screw dislocation under the action of longitudinal shear, the movement and pile-up are strictly along the x-direction, and the solution is exact.